\documentclass[aip,jcp,superscriptaddress,floatfix,preprint]{revtex4-1}
\usepackage{graphicx}
\usepackage{epsfig}
\usepackage{amsmath}
\usepackage{amssymb}
\usepackage{dcolumn}

\newcolumntype{d}[1]{D{.}{.}{#1} }
\newcommand{\la}{\langle}
\newcommand{\ra}{\rangle}
\newcommand{\bea}{\begin{eqnarray}}
\newcommand{\eea}{\end{eqnarray}}

\newcommand{\ch}{\cal H}

\newcommand{\chp}{{\cal H}_p}
\newcommand{\eq}[1]{Eq.~(\ref{#1})}
\newcommand{\fig}[1]{Fig.~\ref{#1}}

\newcommand{\rr}{\mathbf{r}}

\newcommand*{\id}{{\normalfont\hbox{1\kern-0.15em \vrule width .8pt depth-.5pt}}}
\newcommand{\sinv}{(S^{-1})}
\newcommand{\thop}{\hat{\theta}_p}

\newcommand{\tilp}{\tilde{\theta}_p}

\newcommand{\Tilp}{\tilde{\Theta}_p}
\newcommand{\idd}{\hat{\id}}

\begin{document}
\title{An efficient implementation of the localized operator partitioning method for electronic energy transfer}
\author{Jayashree Nagesh}
\affiliation{Chemical Physics Theory Group, Department of Chemistry, University of Toronto, Toronto, ON M5S 3H6, Canada}
\author{Artur F. Izmaylov}
\affiliation{Chemical Physics Theory Group, Department of Chemistry, University of Toronto, Toronto, ON M5S 3H6, Canada}
\affiliation{Department of Physical and Environmental Sciences, University of Toronto, Scarborough, Toronto, ON M1C 1A4, Canada}
\author{Paul Brumer}
\affiliation{Chemical Physics Theory Group, Department of Chemistry, University of Toronto, Toronto, ON M5S 3H6, Canada}
\date{\today}
\begin{abstract}
The localized operator partitioning method [Y. Khan and P. Brumer, J. Chem. Phys. \textbf{137}, 194112 (2012)]
rigorously defines the electronic energy on any subsystem within a molecule and
gives a precise meaning to the subsystem ground and excited electronic energies,
which is crucial for investigating electronic energy transfer from first principles.  However, an efficient implementation of this approach
has been hindered by complicated one- and two-electron integrals arising in its formulation.
Using a resolution of the identity in the definition of partitioning we reformulate the method in a computationally
efficient manner that involves standard one- and two-electron integrals.
We apply the developed algorithm to the $9-$(($1-$naphthyl)$-$methyl)-anthracene (A1N)
molecule by partitioning A1N into anthracenyl and CH$_2-$naphthyl groups as subsystems,
and examine their electronic energies and populations for several excited states using
Configuration Interaction Singles method.
The implemented approach shows a wide variety of different behaviors amongst the excited electronic states.
\end{abstract}
\maketitle

\section{Introduction}
Understanding molecular Electronic Energy Transfer (EET) is of paramount importance, not only for the technological design of better photodevices, but also to deepen our fundamental understanding of energy flow in molecules.\cite{maykuhneet,*parsoneet} A straightforward, first-principles approach to EET would obtain the time dependent molecular wave function $\Psi(t)$ for the system,  define an operator $\chp$ that represents ``electronic energy in a
subunit $p$ of the molecule", and compute energy partitioning as the matrix element of this operator $E_p(t)=\langle\Psi(t)|\chp|\Psi(t)\rangle$.
The local Hamiltonian $\chp$ was designed in Ref.~\onlinecite{yaser_2012},
 guided by three principles: (1) the subsystem energy $E_p$ must be real,
(2) the energies $E_p$ must reduce to the correct component energies for infinitely separated fragments, and
(3) the operator $\chp$ must be symmetric with respect to electron exchange.
The last condition ensures that $E_p$ written as $E_p={\rm Tr}[\mathcal{D} \chp]$ is invariant with respect to electron exchange,
since the electron density $\mathcal{D} = |\Psi(t)\rangle\langle\Psi(t)|$ is symmetric with respect to electron interchange.
The result allows for the computation of EET dynamics exactly, at all inter-chromophoric distances and for all coupling regimes,
(see, e.g., Ref.~\onlinecite{pach_pb2012} and references therein)
and reduces to the F\"{o}rster~\cite{forster} and Dexter~\cite{dexter} theories in corresponding limits.

\begin{figure}
\caption{Plot of partitioned density of anthracenyl (A) subsystem of A1N at an iso-density value of 0.005 $e/a.u.^3$}
\includegraphics[width=8.5cm,height=7cm]{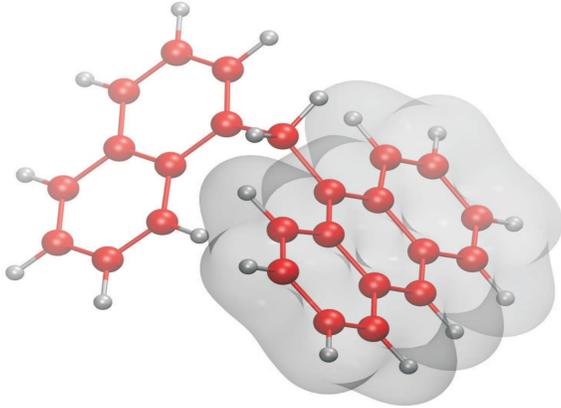}
\label{fig:part_den}
\end{figure}
In spite of many conceptual advantages, the computational algorithm for the localized operator partitioning method (LOPM) 
appears difficult to implement for general molecular systems due to the one- and two-electron integrals over
chromophoric subspaces. This paper introduces an efficient algorithm for partitioning the electronic energies,
allowing the local electronic Hamiltonian to be applied to wave functions of large molecular systems.
The resulting partitioning technique allows a rigorous, quantitative assessment of the amount of energy
delocalization and population distribution in the excited states, from first principles. As a first example, we
consider A1N (Fig.~\ref{fig:part_den}), whose photo-induced EET process has been studied experimentally.\cite{levy_2000}
Specifically we divide A1N into anthracenyl (A) and -CH$_2$-naphthyl (N) subsystems by defining a
partition surface formed by grouping atomic surfaces of the A and N fragments
(see Fig.~\ref{fig:part_den} for the partitioned molecular density obtained using this method),
and apply the LOPM to individual stationary states of the A1N electronic Hamiltonian.

Insofar as the focus here is the development of an efficient tool to evaluate the \textit{partitioning} of the electronic energy, it suffices to consider electronic wave functions resulting from the lowest level non-empirical method for excited states, i.e., Configuration Interaction Singles (CIS). Despite this deliberate simplification, the results are found to provide insight into energy partitioning amongst the set of twelve excited states that were examined. Extensions to higher level theories are beyond the scope of this paper and will be the subject of future work.

\section{Method}\label{sec:loc_op_theory}
\subsection{Localized operator partitioning}
Assuming the Born-Oppenheimer separation of electronic and nuclear coordinates, the electronic Hamiltonian ${\ch}$ (in a.u.) for an $N_e$-electron molecular system is given by
\bea
{\ch} &=& \sum_m h(\mathbf{r}_m) + \sum_{m>n} \frac{1}{|\mathbf{r}_{m}-\mathbf{r}_n|} \notag\\
&+& \sum_{\alpha >\beta} \frac{Z_{\alpha}Z_{\beta}}{|\mathbf{R}_{\alpha}-\mathbf{R}_{\beta}|},  \label{eqn:mol_h}\\
h(\mathbf{r}_m) &=& -\frac{1}{2} \nabla_m^2 - \sum_{\alpha} \frac{Z_{\alpha}}{|\mathbf{R}_{\alpha}-\mathbf{r}_m|},
\eea
where $\mathbf{r}_m$,$\mathbf{r}_n$ and $\mathbf{R}_{\alpha}$, $\mathbf{R}_{\beta}$ are electronic and nuclear coordinates respectively, $\nabla_m^2$ is a
Laplacian in electronic Cartesian components, and $Z_{\alpha}$, $Z_{\beta}$ are nuclear charges.
The nuclear-nuclear repulsion term in Eq.~(\ref{eqn:mol_h}) is a constant since nuclear coordinates are treated as parameters in $\ch$, and therefore
will be neglected below.

For stationary states of ${\ch}$, the local electronic Hamiltonian $\chp$ of subsystem $p$ is given by\cite{yaser_2012}
\bea
\chp &=&  \sum_m \theta_p(\mathbf{r}_m)h(\rr_m)
+\frac{1}{2}\sum_{m\neq n} \frac{\Theta_{p}(\mathbf{r}_m,\mathbf{r}_n)}{|\mathbf{r}_{m}-\mathbf{r}_n|},
\eea
where
\bea
\theta_p(\mathbf{r}_m) &=& \left\{ \begin{array}{ll} 1 & \mbox{if } \mathbf{r}_m \in p \\ 0 & \mbox{otherwise} \end{array} \right.
\eea
is the one-electron projection operator for subsystem $p$, and
\bea
\Theta_{p}(\mathbf{r}_m,\mathbf{r}_n) &=& \frac{1}{2}\sum_q \big[\theta_p(\mathbf{r}_m)\theta_q(\mathbf{r}_n)+\theta_p(\mathbf{r}_n)\theta_q(\mathbf{r}_m)\big] \label{eqn:proj_2e}
\eea
is the two-electron projection operator on subsystems $p$ and $q$, which is symmetric with respect to electron exchange. For electronic eigenstates $\Psi_I$ of ${\ch}$, subsystem energies are given by $E_p^{(I)}=\la\Psi_I|\chp|\Psi_I\ra={\rm Tr}[\Gamma_I \chp]$, where $\Gamma_I$ is the corresponding two-electron density.
Although Eq.~(\ref{eqn:proj_2e}) is the proposed form of $\chp$ in Ref.~\onlinecite{yaser_2012} , since the
two-electron density $\Gamma_I$ is symmetric with respect to the electron coordinate exchange,
$E_p^{(I)}$ is the same whether symmetric or non-symmetric (e.g., $\tilde{\Theta}_{p}(\mathbf{r}_m,\mathbf{r}_n) = \sum_q \theta_p(\rr_m)\theta_q(\rr_n)$)
forms of the two-electron partition operator are used.
Similarly, we define average subsystem electron populations for each electronic state $I$ as
\bea\label{popp}
{\cal N}_p^{(I)} = \la\Psi_I| \sum_m \theta_p(\mathbf{r}_m)|\Psi_I\ra.
\eea

Owing to the completeness of one- and  two-electron projection operators\cite{yaser_2012}
\begin{equation}
\label{locpop}
\sum_p \theta_p (\rr) = \mathbf{1}_\rr, \quad \sum_{p}\Theta_{p}(\rr,\rr') = \mathbf{1}_{\rr,\rr'},
\end{equation}
the subsystem properties $E_p^{(I)}$ and ${\cal N}_p^{(I)}$ are additive: appropriate sums over subsystems reproduce the full system properties.

To obtain the partitioned analogs of electronic energies using finite Gaussian basis sets ($\phi_\mu, \phi_\nu,...$),
one needs to partition the corresponding atomic integrals as
\bea
S_{\mu\nu}^{(p)} &=& \int d\rr \phi_{\mu}(\rr)\theta_p(\rr) \phi_{\nu}(\rr),\\
h_{\mu\nu}^{(p)} &=& \int d\rr \phi_{\mu}(\rr) \theta_p(\rr) h(\rr) \phi_{\nu}(\rr), \label{hp}\\
g_{\mu\nu\lambda\sigma}^{(p)} &=& \int d\mathbf{r}_1 d\mathbf{r}_2
\phi_{\mu}(\mathbf{r}_1)\phi_{\nu}(\mathbf{r}_1) \notag \\
&\times&\frac{\Theta_{p} (\rr_1,\rr_2)}{|\mathbf{r}_1-\mathbf{r}_2|}  \phi_{\lambda}(\mathbf{r}_2)\phi_{\sigma}(\mathbf{r}_2).
\label{eqn:2e_part}
\eea
Although partitioning the kinetic energy and overlap integrals is straightforward,
partitioning the nuclear attraction and electron-electron repulsion integrals involves
modification of the Boys integral\cite{shavitt_1963} to accommodate the altered shape of the partitioned atomic orbitals.
To circumvent the problem of partitioned integral evaluation [Eqs.~(\ref{hp}) and (\ref{eqn:2e_part})],
we redefine the localized operator partitioning by projecting partitioned operators
onto a finite one-electron AO basis. The projection operator is defined as a resolution of the identity (RI)
\begin{equation}
\hat{\id} = \sum_{\mu\nu} |\mu\rangle \sinv_{\nu\mu} \langle \nu|, \label{eqn:ri1e}
\end{equation}
where $\sinv_{\nu\mu}$ are matrix elements of the inverse of the AO overlap matrix $S$
and $|\mu\rangle, |\nu\rangle$ are AO basis functions.
Then, a one-electron projected local operator is
 \bea\label{eq:tilp}
 \tilp &=& \hat{\id} ~\thop \hat{\id} \\
 &=& \sum_{\mu\nu} |\mu\ra L_{\mu\nu}^{(p)} \la \nu|,
\eea
where
\bea
L_{\mu\nu}^{(p)} &=& \sum_{\mu_1\nu_1} \sinv_{\mu \mu_1} S^{(p)}_{\mu_1 \nu_1} \sinv_{\nu_1 \nu}, \\
S^{(p)}_{\mu_1 \nu_1} &=& \la \mu_1 | \thop | \nu_1\ra \\
&=& \int d\rr \phi_{\mu}(\rr)\theta_{p}(\rr)\phi_{\nu}(\rr).
\eea

In contrast to the $\thop$ operator, the $\tilp$ operator is an integral operator
in the coordinate representation
\bea
\la\rr|\tilp|f\ra &=& \int d\rr' \la \rr|\tilp|\rr'\ra \la\rr' | f\ra \\
&=& \sum_{\mu\nu}L_{\mu\nu}^{(p)}\la\rr|\mu\ra \int d\rr' \la\nu | \rr'\ra \la\rr' | f\ra \\
&=& \sum_{\mu\nu}L_{\mu\nu}^{(p)}\phi_{\mu}(\rr) \int d\rr' \phi_{\nu}^{*}(\rr') f(\rr'),
\eea
where $f(\rr')=\la \rr'|f\ra$ is an arbitrary one-electron function.
For a one-electron operator $\hat h$, matrix elements 
\bea\label{eq:hpt}
\la \mu |\hat h\tilp |\nu \ra =  \sum_{\lambda\sigma} \la \mu |\hat h|\lambda \ra S_{\lambda\sigma}^{-1}  S_{\sigma\nu}^{(p)}
\eea
and $\la \mu| \hat h\thop|\nu \ra = h_{\mu\nu}^{(p)}$ are different due to basis set incompleteness.
This is not an issue: we adopt $\tilp$ as our primary partitioning operator.
This projected partitioning is more convenient
in implementation and  gives exactly the same partitioned densities as a non-projected version.
To illustrate this property,  consider the ground state electronic population of subsystem $p$, ${\cal N}_p^{(0)}$, using $ \int d\rr
 \la \rr|\hat{\rho}^{(0)}\hat{\theta}_p|\rr\ra$ and  $\int d\rr \la \rr|\hat{\rho}^{(0)}\tilde{\theta}_p|\rr \ra $,
 where $\hat{\rho}^{(0)}$ is the closed shell ground state density operator,
\begin{equation}
\hat{\rho}^{(0)} = 2\sum_{\mu\nu} |\mu\ra P_{\mu\nu}^{(0)}\la \nu|. \label{eqn:gsdenop}
\end{equation}
When using $\hat{\theta}_p$ we obtain ${\cal N}_p^{(0)} = 2\sum_{\mu\nu} P_{\mu\nu}^{(0)} S_{\mu\nu}^{(p)}$,
whereas with $\tilde{\theta}_p$ we have
\bea
{\cal N}_p^{(0)} &=& 2\sum_{\mu\nu\lambda\sigma} P_{\mu\lambda}^{(0)} S^{(p)}_{\lambda\sigma} \sinv_{\sigma\nu} S_{\mu\nu}\\
&=& 2\sum_{\mu\nu} P_{\mu\nu}^{(0)} S_{\mu\nu}^{(p)}.
\eea
It is convenient to introduce a partitioned one-electron density in the form
\bea
\tilde{P}_{\mu\nu}^{(0,p)}  = \sum_{\lambda\sigma} P_{\mu\lambda}^{(0)} S^{(p)}_{\lambda\sigma} \sinv_{\sigma\nu}
\eea
so that ${\cal N}_p^{(0)}$ can be written as ${\cal N}_p^{(0)} = 2\sum_{\mu\nu} \tilde{P}_{\mu\nu}^{(0,p)} S_{\mu\nu}$.
Figure~\ref{fig:part_den} of the partitioned density of the anthracenyl subsystem (A),
$\rho_0^{(A)}(\rr)=2\sum_{\mu\nu}\tilde{P}_{\mu\nu}^{(0,A)}\phi_{\mu}(\rr)\phi_{\nu}(\rr)$, illustrates
that the partitioned density is indeed localized on the anthracenyl portion of A1N. 

Since the symmetrization in the two-electron part of the partitioned Hamiltonian
is redundant we introduce the non-symmetric two-electron projected operator
\begin{equation}
\Tilp = \tilp\otimes\idd = \sum_{\mu\nu\lambda\sigma}|\mu\lambda\ra L_{\mu\nu}^{(p)}L_{\lambda\sigma}\la\nu\sigma|,
\end{equation}
where $|\mu\lambda\ra$ and $|\nu\sigma\ra$ are two-electron AO basis kets and $L_{\lambda\sigma}=\sum_q L_{\lambda\sigma}^{(q)}$.
The partitioned two-electron integrals become
\bea
\tilde{g}_{\mu\nu\lambda\sigma}^{(p)} &=& \sum_{\mu_1\nu_1\lambda_1\sigma_1} S_{\mu\mu_1}^{(p)}S_{\lambda\lambda_1}\sinv_{\mu_1\nu_1}\sinv_{\lambda_1\sigma_1}\la\nu_1\sigma_1|\nu\sigma\ra \notag \\
&=& \sum_{\mu_1\nu_1} S_{\mu\mu_1}^{(p)}\sinv_{\mu_1\nu_1}\la\nu_1\lambda|\nu\sigma\ra,\label{eqn:gpart}
\eea
where
\bea
\la \nu_1 \lambda|\nu\sigma\ra = \int d\mathbf{r}_1 d\mathbf{r}_2 \frac{\phi_{\nu}(\mathbf{r}_1)\phi_{\nu_1}(\mathbf{r}_1)\phi_{\sigma}(\mathbf{r}_2)\phi_{\lambda}(\mathbf{r}_2)}{|\mathbf{r}_1-\mathbf{r}_2|},\label{eqn:2e}
\eea
are regular two-electron integrals in the Dirac notation.  Thus, the projecting partitioning operator $\tilde{\theta}_p$ [\eq{eq:tilp}] allows us to use 
standard integrals in both one- and two-electron contributions, Eqs.~(\ref{eq:hpt}) and (\ref{eqn:gpart}), and therefore by-passes the problem 
of partitioning the Boys integral.  

\subsection{Partitioning in Configuration Interaction Singles (CIS) method}
The CIS wavefunction for the $I^{th}$ singlet state is $\Psi_I = \sum_{ia} c_{ia}^{(I)} (\Phi_{ia}^{\alpha\beta}-\Phi_{ia}^{\beta\alpha})/\sqrt{2}$,
where $\Phi_{ia}^{\alpha\beta}$ is a Slater determinant obtained from the Hartree-Fock (HF) determinant $\Psi_0$ by replacing a spin molecular orbital (MO) $\phi_i^{\alpha}$ by virtual MO $\phi_a^{\beta}$. The CIS energy $E^{(I)}$ is determined as the sum of HF ($E^{(0)}$) and excitation ($\Delta E^{(I)}$) energies. The $E^{(0)}$ in terms of spatial atomic orbitals (AOs) $\phi_{\mu},\phi_{\nu}\ldots$ is given by
\bea
E^{(0)} &=& 2\sum_{\mu\nu} P_{\mu\nu}^{(0)} h_{\mu\nu} \nonumber \\
&+& \sum_{\mu\nu\lambda\sigma}[2P_{\mu\nu}^{(0)}P_{\lambda\sigma}^{(0)} - P_{\mu\sigma}^{(0)}P_{\lambda\nu}^{(0)}]\la \mu\lambda|\nu\sigma\ra, \label{eqn:ehf}
\eea
where the HF electron density $P^{(0)}_{\mu\nu} = \sum_{i=1}^{N_e/2} C_{i\mu}C_{i\nu}$,
obtained from  the MO coefficients $C_{i\mu}$ and $C_{i\nu}$ assuming the closed-shell ground state.
Throughout this paper $a,b,...$ subscripts are used for the HF virtual and $i,j,...$ subscripts for the HF occupied spatial MOs.
The excitation part of the electron energy in the AO representation is
\bea
\Delta E^{(I)} &=& 2\sum_{\mu\nu} R_{\mu\nu}^{(I)} h_{\mu\nu} +\sum_{\mu\nu\lambda\sigma} \Gamma_{\mu\nu\lambda\sigma}^{(I)} \la\mu\lambda|\nu\sigma\ra, \label{eqn:cis_full}
\eea
where
\bea
\Gamma_{\mu\nu\lambda\sigma}^{(I)} &=& 2P_{\mu\nu}^{(0)}R_{\lambda\sigma}^{(I)}-P_{\mu\sigma}^{(0)}R_{\lambda\nu}^{(I)}+2R_{\mu\nu}^{(I)}P_{\lambda\sigma}^{(0)}-R_{\mu\sigma}^{(I)}P_{\lambda\nu}^{(0)} \notag \\
&+& 2T_{\mu\nu}^{(I)}T_{\sigma\lambda}^{(I)} - T_{\mu\sigma}^{(I)}T_{\nu\lambda}^{(I)}+ 2T_{\nu\mu}^{(I)}T_{\lambda\sigma}^{(I)} - T_{\sigma\mu}^{(I)}T_{\lambda\nu}^{(I)},\notag\\
R_{\mu\nu}^{(I)} &=& \sum_{ijab}c_{ia}^{(I)} c_{jb}^{(I)} \big\{ \delta_{ij}C_{b\mu}C_{a\nu} - \delta_{ab}C_{i\mu}C_{j\nu} \big\},\\
T_{\mu\nu}^{(I)} &=& \sum_{ia}c_{ia}^{(I)} C_{i\mu}C_{a\nu}.
\eea
The one-electron contributions are evaluated by replacing $\hat h$ with $\hat h \tilp$ in one-electron integrals
for the ground state energy
\bea
\sum_{\mu\nu} P_{\mu\nu}^{(0)} \la\mu|\tilp \hat{h}|\nu\ra &=& \sum_{\mu\nu\lambda\sigma} P_{\mu\nu}^{(0)} S_{\mu\lambda}^{(p)} \sinv_{\lambda\sigma} h_{\sigma\nu} \nonumber  \\
&=& \sum_{\nu\sigma} \tilde{P}_{\nu\sigma}^{(0,p)} h_{\sigma\nu} \label{eqn:part1_gs}
\eea
and for the excitation energy
\bea
\sum_{\mu\nu} R_{\mu\nu}^{(I)} \la \mu|\tilp \hat{h}|\nu\ra &=& \sum_{\nu\sigma} \tilde{R}_{\nu\sigma}^{(I,p)} h_{\sigma\nu}, \label{eqn:part1_es} \\
\tilde{R}_{\nu\sigma}^{(I,p)} &=& \sum_{\mu\lambda} R_{\mu\nu}^{(I)} S_{\mu\lambda}^{(p)} \sinv_{\lambda\sigma}. \label{eqn:qden}
\eea
The two-electron contribution to the partitioned ground state energy is given by
\bea
&& \sum_{\mu\nu\lambda\sigma}[2P_{\mu\nu}^{(0)}P_{\lambda\sigma}^{(0)} - P_{\mu\sigma}^{(0)}P_{\lambda\nu}^{(0)}]\tilde{g}_{\mu\nu\lambda\sigma}^{(p)}\nonumber\\
&=& \sum_{\mu\nu\lambda\sigma}[2P_{\mu\nu}^{(0)}P_{\lambda\sigma}^{(0)} - P_{\mu\sigma}^{(0)}P_{\lambda\nu}^{(0)}] \sum_{\mu_1\nu_1} S_{\mu\mu_1}^{(p)}\sinv_{\mu_1\nu_1}\la\nu_1\lambda|\nu\sigma\ra \nonumber \\
&=& \sum_{\nu_1\lambda\nu\sigma} \Gamma_{\nu\nu_1\sigma\lambda}^{(0,p)}\la\nu_1\lambda|\nu\sigma\ra,
\eea
where
\bea
\Gamma_{\nu\nu_1\sigma\lambda}^{(0,p)} = 2\tilde{P}_{\nu\nu_1}^{(0,p)}P_{\lambda\sigma}^{(0)} - \tilde{P}_{\sigma\nu_1}^{(0,p)}P_{\lambda\nu}^{(0)}.
\eea
Similar partitioning of the excited two-electron density matrix $\Gamma_{\mu\nu\lambda\sigma}^{(I)}$ leads to
\bea
\Gamma_{\mu\nu\lambda\sigma}^{(I,p)} &=& 2\tilde{P}_{\mu\nu}^{(0,p)}R_{\lambda\sigma}^{(I)} - \tilde{P}_{\mu\sigma}^{(0,p)}R_{\lambda\nu}^{(I)}+\tilde{R}_{\mu\nu}^{(I,p)}P_{\lambda\sigma}^{(0)} - \tilde{R}_{\mu\sigma}^{(I,p)}P_{\lambda\nu}^{(0)} \notag\\
&+& 2\tilde{T}_{\mu\nu}^{(I,p)}T_{\sigma\lambda}^{(I)} - T_{\mu\sigma}^{(I,p)}T_{\nu\lambda}^{(I)} + 2\tilde{T}_{\nu\mu}^{(I,p)}T_{\lambda\sigma}^{(I)}-\tilde{T}_{\sigma\mu}^{(I,p)}T_{\lambda\nu}^{(I)}\notag\\
\eea
where $\tilde{T}_{\mu\nu}^{(I,p)} = \sum_{\lambda\sigma} T_{\mu\lambda}^{(I)} S_{\lambda\sigma}^{(p)} \sinv_{\sigma\nu}$.
Using the partitioned densities $\tilde{P}^{(0,p)}$, $\tilde{R}^{(I,p)}$, $\Gamma^{(0,p)}$ and $\Gamma^{(I,p)}$ we write the partitioned energies within the CIS theory as
\bea
E_p^{(I)} &=& E_p^{(0)} + 2\sum_{\mu\nu} \tilde{R}_{\mu\nu}^{(I,p)} h_{\mu\nu} + \sum_{\mu\nu\lambda\sigma} \Gamma_{\mu\nu\lambda\sigma}^{(I,p)} \la\mu\lambda|\nu\sigma\ra, \notag \\
\label{eqn:espden}
\eea
where
\bea
E_p^{(0)} &=& 2\sum_{\mu\nu} \tilde{P}_{\mu\nu}^{(0,p)} h_{\mu\nu} + \sum_{\mu\nu\lambda\sigma} \Gamma_{\mu\nu\lambda\sigma}^{(0,p)}\la\mu\lambda|\nu\sigma\ra. \label{eqn:gspden}
\eea
Thus, using the projected localized partitioning allows us to formulate all computationally intense partitioned
quantities as a product of standard Gaussian integrals contracted with various densities.
In these contractions we use earlier developed screening and fast multipole moment techniques.\cite{ahlrichs_1974,Scuseria:1999/JPC/4782,Izmaylov:2006/JCP/104103}

\section{Results and discussion}
Calculations on A1N were carried out within the Gaussian suite of programs \cite{g09} using CIS/6-31G level of theory at the HF optimized geometry of the ground state.
A1N is partitioned into anthracenyl (A) and -CH$_2$-naphthyl (N) subsystems based on atom grouping: $\Omega_A$ ($\Omega_N$) is a set of all atoms in the A (N) fragment.
To define each atom and the collective separating surface
we used Becke's partitioning scheme,\cite{becke88}
and the partitioned overlaps are obtained as weighted sums
\bea
S_{\mu\nu}^{(A)} &=& \sum_{k\in\Omega_A} \sum_i w(\rr_i) p_k(\rr_i) \phi_{\mu}(\rr_i)\phi_{\nu}(\rr_i),
\eea
where $w(\rr_i)$ is the quadrature weight associated with grid point $\rr_i$, and $p_k(\rr_i)$ is the atomic partition function.\cite{becke88} A spherical quadrature scheme consisting of a product of Mura-Knowles\cite{mura-knowles} radial and Murray-Handy-Laming\cite{mhl} angular quadratures is used to evaluate atomic contributions to the overlap integrals.\cite{note:numdet}

\begin{table}[!h]
\caption{Local populations and excitation energies; $\Delta {\cal N}_p^{(I)} = {\cal N}_p^{(I)} - {\cal N}_p^{(0)}$
and $\Delta E_p^{(I)} = E_p^{(I)} - E_p^{(0)}$, where $p=$A,N; $\Delta E^{(I)} = E^{(I)} - E^{(0)}$;
$\sum_p {\cal N}_p^{(I)} = N_e$ for all $I$; In A1N, $N_e=168$ and $\mathrm{{\cal N}^{(0)}_A}\approx 92.93$.
The type of each excited state is assigned based on
qualitative analysis of electron density distributions corresponding to occupied-virtual (OV) MO pairs with
large CIS coefficients ($|c_{ia}^{(I)}|>0.1$).
}\label{tab:results}
\begin{tabular}{c c c c c}
\hline
\hline
State $I$ & \multicolumn{1}{c}{$\mathrm{\Delta{\cal N}_A^{(I)}}$ } & \multicolumn{1}{c}{$\Delta E_{\mathrm{A}}^{(I)}$/$\Delta E^{(I)}$} & \multicolumn{1}{c}{$\Delta E_{\mathrm{N}}^{(I)}$/$\Delta E^{(I)}$} & Type \\
\hline
    1 &   0.00  &  1.40  & -0.40 & Local on A\\
    2 &   0.00  &  1.51  & -0.51 & Local on A \\
    3 &   0.00  & -0.03  &  1.03 & Local on N \\
    4 &   0.00  & -0.19  &  1.19 & Local on N \\
    5 &   0.01  &  0.72  &  0.28 & Local on A \\
    6 &   0.56  &-22.32  & 23.32 & CT: N $\rightarrow$ A\\
    7 &   0.03  & -0.05  &  1.05 & Local on A\\
    8 &  -0.56  & 22.82  &-21.82 & CT: A $\rightarrow$ N\\
    9 &  -0.03  &  2.40  & -1.40 & Local on A\\
   10 &  -0.01  &  1.40  & -0.40 & Local on A\\
   11 &   0.00  &  1.56  & -0.56 & Local on A\\
   12 &  -0.01  &  0.26  &  0.74 & Local on N\\
\hline
\hline
\end{tabular}
\end{table}

The results of applying the LOPM to the electronic populations (${\cal N}_p^{(I)}$) and energies ($E_p^{(I)}$)
of the two subsystems in the first 12 singlet excited states of A1N are presented in Table \ref{tab:results}.
The magnitude of the ratios of subsystem excitation energies ($\Delta E_p^{(I)}$) with respect to system
excitation energies ($\Delta E^{(I)}$) reflects the extent of localization of excitation energy on the subsystems.
The electronic population change $\Delta{\cal N}_A^{(I)}$ shows changes in charge distribution associated  with the excited state charge density.

\begin{figure}
\caption{CIS/6-31G excitation energies $\Delta E^{(I)}$ (eV) of A1N, anthracene (Anth) and methyl-naphthalene (MN). A1N excited states are color coded based on extent of localization ($\Delta E_p^{(I)}$ magnitude): blue (dashed) for the anthracenyl part, red (dotted) for the -CH$_2$-naphthyl part, and black (solid) for delocalized states.}
\includegraphics[width=8.5cm,height=10.5cm]{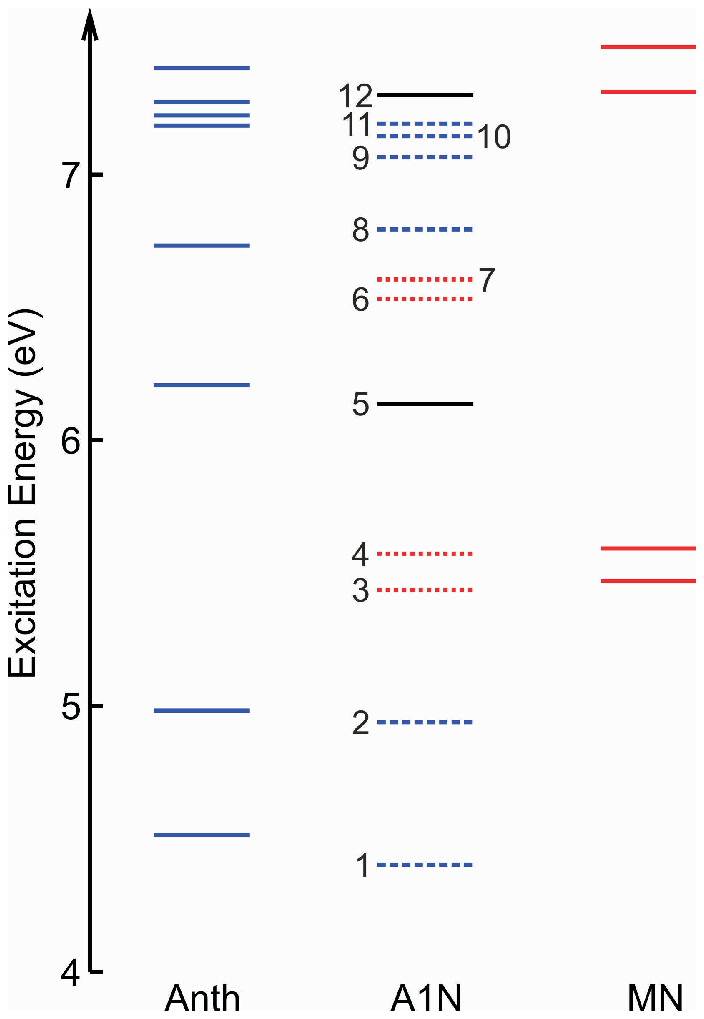}
\label{fig:a1n_levels}
\end{figure}

\textit{Chromophoric States}.  Consider first states 1-4,9-11. In this scheme, a positive value of $\Delta E_p^{(I)}/\Delta E^{(I)}$
indicates that the electronic excitation energy is localized on fragment $p$: in states 1,2,9-11, it is anthracenyl subsystem, and in states 3 and 4, it is the -CH$_2$-naphthyl subsystem.
The localization of excitation in states 1-4,9-11 is consistent with the qualitative picture that emerges from analyzing molecular orbitals involved in these excitations.
We also note that a comparison of excitation energies of localized states 1-4,9-11 with those of isolated molecules in Fig.~\ref{fig:a1n_levels} at the same geometries as in A1N (Fig.~\ref{fig:part_den}) is excellent: states 3,4 with methyl-naphthalene and states 1,2,9-11 with anthracene. The agreement of energies in Fig.~\ref{fig:a1n_levels} along with negligible change in subsystem populations in Table~\ref{tab:results} suggests that these excited electronic states are well described as localized excitations on chromophores.

\begin{figure}
\caption{Orbital energy diagram illustrating occurrence of negative partitioning energy differences: (left) occupied ($i$) and virtual ($a$) orbitals
are both localized on fragment N; (right) occupied ($i$) and virtual ($a$) orbitals are localized on fragments N and A, respectively.}
\includegraphics[width=0.45\textwidth]{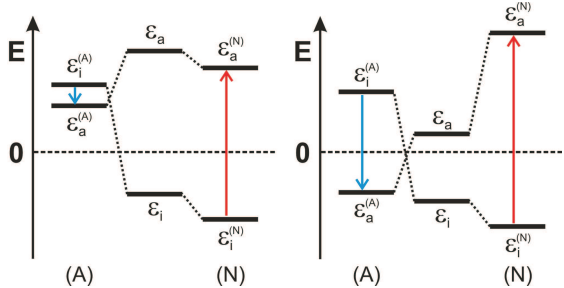}
\label{fig:ediag}
\end{figure}

\textit{Negative Ratios}.  Partitioning of energy, as shown in Table I, has been referred to the ground state energies.  
The alternative, absolute numbers associated with each state, are not used here because the energy 
partitioning would be strongly biased by the number of electrons in each partition.  
However, referencing results to the ground state has the interesting feature of the appearance of de-excitation or negative ratios
$\Delta E_p^{(I)}/\Delta E^{(I)}$ in states with localized or charge transfer (CT) excitations (e.g., state 4, state 6). Since the total electronic
wave-functions $\Psi_I$ are variationally optimized for the total electronic Hamiltonian rather than for its partitioned
counterpart, $\Delta E_p^{(I)} = E_p^{(I)} - E_p^{(0)}$ can be negative. To understand further how this negativity arises,
 consider a simplified case when the excited state is obtained by substituting a single occupied MO $\phi_i$
with a single unoccupied MO $\phi_a$. In this case, the full excitation energy is given by
$\Delta E^{I}=\varepsilon_a-\varepsilon_i$, where $\varepsilon_i$ and $\varepsilon_a$ are corresponding MO energies.
These energies are
\begin{eqnarray}
\varepsilon_t &=& h(\rho_t) + J(\rho_t,\rho_0) + K(\rho_t,\rho_0),  \label{eqn:orb_full}
\eea
where $t=a$ and $i$,
\bea
h(\rho_t) &=& \int d\rr [h(\rr) \rho_t(\rr,\rr')]_{\rr'=\rr}, 
\eea
\bea
J(\rho_t,\rho_0) &=& \int d\rr d\rr' \frac{\rho_t(\rr,\rr)\rho_0(\rr',\rr')}{|\rr-\rr'|}, 
\eea
\bea
K(\rho_t,\rho_0) &=& \int d\rr d\rr' \frac{\rho_t(\rr,\rr')\rho_0(\rr',\rr)}{|\rr-\rr'|},
\end{eqnarray}
and $\rho_t(\rr,\rr') = \phi_t(\rr) \phi_t(\rr')$, $\rho(\rr,\rr') = \sum_i \phi_i(\rr) \phi_i(\rr')$ are orbital and
total ground state densities, respectively.
The partitioned excitation energies are $\Delta E_p^{(I)}=\varepsilon_a^{(p)}-\varepsilon_i^{(p)}$, $p=A,N$, and
their negative values arise from the interplay between two
largest contributions in orbital energies: the electron-nuclear ($V_{ne}$)
part of $h_{tt}$ and $J(\rho_t,\rho_0)$ upon their partitioning,
thus we will focus only on these two terms. Usually, negativity appears as a result of partitioned
virtual orbital energy going below the Fermi energy and/or occupied partitioned orbital energy going above the Fermi energy.
These shifts take place when participating orbitals are localized, therefore, for illustration we will consider a partitioned
orbital energy $\varepsilon_t^{(N)}$ with $\rho_t(\rr,\rr')$ localized on the
-CH$_2$-naphthyl fragment, $\rho_t\approx\rho_t^{(N)}$. In this case,
$V_{ne}(\rho_t)^{(N)}$ has almost full negative value of $V_{ne}(\rho_t)$ due to the orbital localization.
On the other hand, the partitioned Coulomb orbital energy is
\bea
J^{(N)}(\rho_t,\rho_0) &=& \frac{1}{2} \left[J(\rho_t^{(N)},\rho_0)+J(\rho_t,\rho_0^{(N)})\right].
\eea
Note that the averaging in $J^{(N)}(\rho_t,\rho_0)$ comes as a result of taking into
account the indistinguishability of electrons.
Applying the orbital localization condition, $\rho_t^{(N)}\approx \rho_t$, $J(\rho_t,\rho_0)^{(N)}$ can be approximated as
\bea
J^{(N)}(\rho_t,\rho_0) &\approx& \frac{1}{2} \left[J(\rho_t,\rho_0)+J(\rho_t,\rho_0^{(N)}) \right].
\eea
Considering the ratio between the number of electrons on the -CH$_2$-naphthyl fragment and the total number of electrons as
an estimate to the ratio $J(\rho_t,\rho_0^{(N)})/J(\rho_t,\rho_0)$, $J(\rho_t,\rho_0)^{(N)}$ becomes
\bea
J(^{(N)}\rho_t,\rho_0) &\approx& \frac{1}{2} \left[J(\rho_t,\rho_0)+  \frac{76}{168} J(\rho_t,\rho_0)\right] \\
&\approx& \frac{3}{4}J(\rho_t,\rho_0).
\eea
This result for $J^{(N)}(\rho_t,\rho_0)$  can lead to negative $\varepsilon_t^{(N)}$
because $\varepsilon_t^{(N)}$'s negative component $V_{ne}^{(N)}(\rho_t)$
preserves the full value of its unpartitioned counterpart $V_{ne}(\rho_t)$, while
$\varepsilon_t^{(N)}$'s positive Coulomb contribution
has been reduced by a quarter.
As a result, partitioned orbital energies $\varepsilon_t^{(N)}$ for orbitals localized in the $N$ subsystem
are lower in energy than their unpartitioned counterparts $\varepsilon_t$ [see \fig{fig:ediag} (left)].
For the fragment $A$, the $\varepsilon_t^{(A)}$ becomes positive because of the opposite effect: reduction
of the negative contribution, $V_{ne}^{(A)}(\rho_t)\approx 0$, and non-negligible positive
Coulomb contribution, $J^{(A)}(\rho_t,\rho_0)\approx J(\rho_t,\rho_0)/4$ [\fig{fig:ediag} (left)].
To understand how negativity arises in $\varepsilon_a^{(A)}-\varepsilon_i^{(A)}$ we need to note
that localization of MOs is always partial and virtual orbitals are generally more delocalized
than the occupied ones. Larger delocalization of virtual orbitals leads to the negative contribution
$V_{ne}(\rho_a^{(A)})-V_{ne}(\rho_i^{(A)})<0$ that outweights $J^{(A)}(\rho_a,\rho_0)-J^{(A)}(\rho_i,\rho_0)>0$
resulting in $\varepsilon_a^{(A)}-\varepsilon_i^{(A)}<0$, as is the case for states 3 and 4 in Table I.
The same relations, up to exchange of partitioning labels $N$ and $A$, hold for
states 1,2,9-11 where occupied and virtual orbital densities localized on the $A$ fragment.
Of course, if both the occupied and virtual
orbitals are completely localized on the same fragment, both the partitioned orbital energies experience similar increase or decrease and the partitioned excitation
energies are positive. Due to electron density delocalization the positivity of $\Delta E^{(p)}$ for all $p$ can only be guaranteed at large separation between fragments.

The magnitude of negative partitioned excitation energies becomes much larger
in states with significant CT excitations (states 6 and 8).
In CT excitations, there are usually well defined dominant pairs of
orbital energies ($\varepsilon_a-\varepsilon_i$) contributing to the excitation energy,
both of the involved MOs being localized on different fragments.
For example, in state 6, the dominant excitation is
from an occupied MO localized on -CH$_2$-naphthyl to a virtual MO localized on the anthracenyl subsystem.
The orbital energy analysis presented above reveals for the CT case, that an
increase in $\varepsilon_a^{(p)}$ is accompanied by a decrease in $\varepsilon_i^{(p)}$
and vice versa [see~\fig{fig:ediag} (right)].
This leads to very large differences in partitioned excitation energies for the fragments.

\textit{CIS Coefficients}.  Comparing LOPM results to that anticipated from the CIS coefficients shows some interesting differences.  For example, in state 7, partitioning of energy yields localization of excitation on the -CH$_2$-naphthyl subsystem,
whereas a qualitative analysis based on predominant CIS coefficients predicts excitation energy
localization on the anthracenyl fragment.
To understand this difference we analyzed main orbital excitations contributing to state 7.
Predominant CIS configurations correspond to three types of excitations:
CT: $N\rightarrow A$, CT: $A\rightarrow N$, and local $A\rightarrow A$.
The negative excitation energy, $\varepsilon_a^{(A)}-\varepsilon_i^{(A)}$, due to CT: $N\rightarrow A$
[\fig{fig:ediag} (right)] is compensated by the positive
excitation energy due to local $A\rightarrow A$ and CT: $A\rightarrow N$ excitations.
Thus, generally, when a state is comprised of more than one type of excitation,
the LOPM does not necessarily correlate with the qualitative
analysis based on predominant CIS coefficients.

In states 5 and 12 the predominant CIS coefficients correspond to excitations localized on fragments with some CT character.
For these states even if there is reordering of partitioned orbitals, the CT CI coefficients are small enough that they do not result in negativity
of the overall subsystem excitation.
Comparison of the excitation energy of state 5 with that of the isolated species in Fig.~\ref{fig:a1n_levels}
 reveals only one nearby anthracene level, and no methyl-naphthalene levels.
The lack of a methyl-naphthalene level is also true for states 6 and 7, whose energies are localized in the -CH$_2$-naphthyl subsystem.
This is not surprising, since A1N is a bound system of anthracenyl and -CH$_2$-naphthyl radicals,
formed by a complex balance of attractive (nuclear-electronic and exchange) and repulsive
(inter-electronic Coulomb) interactions among all electronic states of the isolated species.
Hence, analyzing the delocalized A1N excited states in terms of nearby isolated excited states is not necessarily sensible.
Similarly, analyzing A1N in terms of anthracenyl and methyl-naphthyl radicals, the result of separating the
A1N system to infinity, would require an enormous number of subsystem basis states.

\textit{Applicability to Electronic Energy Transfer}.  The above results show that the LOPM provides a clear advantage for analyzing
excited states involved in EET in terms of the chromophores without seeking a separated molecular basis.
As such, the approach is completely general, and applicable, for example, to the chromophoric units in chains
of organic polymers, which are subject to frequent $\pi-\pi$ stacking and where  F\"{o}rster theory fails.\cite{kee}
As subsystem units (intra- or inter-molecular) separate, the CT character of the states decreases,
and the magnitude of negative excitation ratios on a subsystem also decreases.
Hence the LOPM also provides an exact means to study CT-dominant states,
and the possible role of CT excitations in EET.

The reported subsystem electronic excitation energies and populations are
at the geometry corresponding to the minimum of HF ground state energy. As a next step,
the LOPM will be applied to studying dynamics of EET process in A1N starting
with vertical excitation at the ground state minimum to the lowest excited state with
excitation localized on the -CH$_2$-naphthyl subsystem, state 3 in Table~\ref{tab:results}.
Here, as a trial calculation we  optimized the geometry of the 3rd excited state and obtained
$\Delta E_A^{(3)}/\Delta E^{(3)} = 1.52$ at the energy
minimum of this state. This result is in sharp contrast with $\Delta E_A^{(3)}/\Delta E^{(3)} = -0.03$ calculated
the ground state minimum (Table~\ref{tab:results}). These single point results indicate EET processes taking place
while the system undergoes rearrangement of nuclear geometry.
The anthracenyl population remains constant at both ground and state 3 minima,
and the excitation energy is transferred from the -CH$_2$-naphthyl subsystem to the anthracenyl subsystem,
which is consistent with A1N experimental data.\cite{levy_2000}
Clearly, the current approach will facilitate understanding EET, by monitoring subsystem
electronic energies in conjunction with nuclear dynamics.

\section{Concluding remarks}
This paper presents an efficient algorithm to applying the localized operator partitioning method (LOPM) for general molecular systems, by a combination of the resolution of the identity approach, analytic integrals, and numerical overlap evaluation schemes. The utility of the algorithm has been demonstrated by evaluating partitioned CIS excitation energies for the A1N molecule, with a wide variety of behaviors observed amongst excited electronic states. Extending the current framework to open-shell systems and other electronic structure methods (e.g., CASSCF) is readily envisioned. A significant extension currently underway incorporates non-stationary electronic and nuclear dynamics to explore photo-induced electronic energy transfer.\cite{levy_2000} Given a total non-stationary electron-nuclear wavefunction resulting, e.g., from radiative excitation, the current formalism will enable a description of electronic energy transfer between subsystems from first principles.

\section*{Acknowledgements}
P.B. and J.N. gratefully acknowledge financial support from the Air Force Office of Scientific Research under Contract No.FA955-13-1-0005.
A.F.I. thanks the Natural Sciences and Engineering Research Council of Canada (NSERC) for financial support through the Discovery Grants Program.


\begin{thebibliography}{17}
\bibitem [{not()}]{note:numdet}%
  \BibitemOpen
  \href@noop {} {}\bibinfo {note} {A spherical grid of 30 radial, 30 polar, and
  60 azimuthal points was found to be sufficient to achieve $10^{-5}$ accuracy
  in energies, and $10^{-7}$ accuracy in overlap integrals.}\BibitemShut
  {Stop}%
\end{thebibliography}
\providecommand{\noopsort}[1]{}\providecommand{\singleletter}[1]{#1}%

\end{document}